\documentclass[10pt,conference]{IEEEtran}

\usepackage[dvipsnames]{xcolor}
\usepackage[T1]{fontenc}
\usepackage[utf8]{inputenc}
\usepackage{booktabs}
\usepackage{graphicx}
\usepackage{array}
\usepackage{balance}
\usepackage[hidelinks]{hyperref}
\usepackage{xcolor}
\usepackage{listings}
\usepackage{upquote}   % straight quotes in verbatim/listings

\definecolor{pykw}{RGB}{0,51,179}      % Python keywords
\definecolor{sqlkw}{RGB}{136,0,112}    % SQL keywords
\definecolor{strc}{RGB}{166,61,0}      % strings
\definecolor{cmt}{RGB}{110,110,110}    % comments
\definecolor{rule}{RGB}{200,200,200}   % frame

\lstdefinestyle{pysql}{
  language=Python,
  deletecomment=[s]{"""}{"""},
  deletestring=[b]",
  literate={"""}{{\textcolor{strc}{"""}}}3,
  morekeywords=[3]{WITH,AS,SELECT,DISTINCT,FROM,WHERE,IS,NOT,NULL,AND,
                   COUNT,JOIN,ON,GROUP,BY,ORDER,DESC},
  keywordstyle=\color{pykw}\bfseries,
  keywordstyle=[3]\color{sqlkw}\bfseries,
  stringstyle=\color{strc},
  commentstyle=\color{cmt}\itshape,
  basicstyle=\ttfamily\footnotesize,
  showstringspaces=false,
  keepspaces=true,
  numbers=left,
  numberstyle=\tiny\color{cmt},
  numbersep=6pt,
  xleftmargin=1.5em,
  frame=tb,
  rulecolor=\color{rule},
  aboveskip=0.5\baselineskip,
  belowskip=0.5\baselineskip,
  columns=flexible,
  breaklines=false, 
}

\makeatletter
\def\lst@makecaption#1#2{%
  \@makecaption{#1}{#2}%
  \lst@IfSubstring t\lst@captionpos{\vskip\belowcaptionskip}{}}
\makeatother
\usepackage{tikz}
\usetikzlibrary{arrows.meta,positioning}
\newcommand{\rqpartlead}[1]{\smallskip\noindent\textbf{#1}}
\usepackage{xcolor} 
\begin{document}

%\title{MCP-Eco: Mining Both Sides of the Model Context Protocol (MCP) Ecosystem}
\title{MCP-Eco: A Linked Dataset of MCP Servers, Consumers, and Development Activity}
\IEEEoverridecommandlockouts

\author{
\IEEEauthorblockN{Khalid Tounoussi\IEEEauthorrefmark{1},
                  Mohammed Sayagh\IEEEauthorrefmark{1},
                  Ali Ouni\IEEEauthorrefmark{1},
                  Yutaro Kashiwa\IEEEauthorrefmark{3}}
\IEEEauthorblockA{\IEEEauthorrefmark{1} \'ETS Montr\'eal, University of Quebec, Montreal, Canada\\
                  khalid.tounoussi.1@ens.etsmtl.ca,
                  \{mohammed.sayagh, ali.ouni\}@etsmtl.ca}
\IEEEauthorblockA{\IEEEauthorrefmark{3}Nara Institute of Science and Technology (NAIST),
                  Nara, Japan\\
                  yutaro.kashiwa@is.naist.jp}
}

\maketitle
\begin{abstract}
The Model Context Protocol (MCP) enables LLM agents to access external tools and data. Studying its software ecosystem requires connecting server interfaces, the projects that configure these servers, and their development activity. However, this information is scattered across directories, package registries, configuration files, and repository histories. We present MCP-Eco, a relational dataset that integrates these sources and links consumer configuration declarations to candidate server projects. Its frozen release contains 16,880,899 records across 19 Parquet tables, covering 78,302 candidate server projects, 16,165 candidate consuming repositories, and 9,200 configuration declarations. Of these declarations, 5,975 resolve to 5,138 distinct repository–server pairs. The dataset also includes tool, resource, and prompt descriptions, declared dependencies, and associated development records, including pull requests, reviews, issues, and commits. Provenance fields distinguish directory-reported capabilities from those observed through live probing and document discovery sources and declaration-matching methods. These connections support research on MCP interface design, configuration practices, and repository maintenance, while enabling practitioners to inspect tool schemas and existing integration configurations. The dataset and accompanying analysis scripts are publicly available at
\url{https://huggingface.co/datasets/mcp-research-alliance/mcp-ecosystem}

Demonstration video:
\url{https://www.youtube.com/watch?v=N7TqnNneSfU}
\end{abstract}


\section{Introduction}
Large language model (LLM) agents act through external tools: they read files, query services, and change their environment. The Model Context Protocol (MCP), introduced by Anthropic in November 2024~\cite{anthropic2024mcp}, standardizes how agents discover and invoke such tools~\cite{hou2025mcp}. An MCP \emph{server} exposes \emph{tools} that perform actions, \emph{resources} that provide data, and \emph{prompts} that supply reusable instructions. A project that wants to use a server declares it in a configuration file, which an agent host such as Claude Desktop or Cursor reads at runtime to connect to each listed server. The ecosystem therefore has two sides: the projects that implement servers and the projects that consume them.

Both sides are software engineering subjects. A server is an ordinary software project with interface design, documentation, dependency, and maintenance concerns, with one difference: an LLM selects among its tools by reading natural-language descriptions and input schemas, so description quality can change agent behavior~\cite{hasan2026smelly}. A consuming project’s configuration records which servers its developers chose and how they launch them. Both sides receive contributions from coding agents such as Claude Code and Copilot.

Several studies and datasets have begun to describe the ecosystem. Hasan et al.\ conducted the first large-scale empirical study of MCP servers, examining the health, security, and maintainability of 1,899 servers~\cite{hasan2025mcpfirstglance}, and later identified quality smells in tool descriptions~\cite{hasan2026smelly}. MCPCorpus~\cite{lin2025mcpcorpus} collects about 14,000 servers and 300 clients with normalized GitHub metadata. MCPZoo~\cite{wu2025mcpzoo} targets runnability, packaging servers so that they can be launched and queried. Guo et al.~\cite{guo2025measurement} measure 8,060 servers and 341 clients.
These resources share three limits. First, their client side consists of host and client implementations, which speak the protocol, rather than the projects that adopt servers through configuration, so none links a server to the repositories that declare it. Second, they capture repository metadata at one point in time but not the pull requests, reviews, issues, and commits through which servers and adopters change. Third, where capability lists are provided, each comes from one channel, either a directory listing or a probe, so advertised and observed interfaces cannot be compared within one resource.

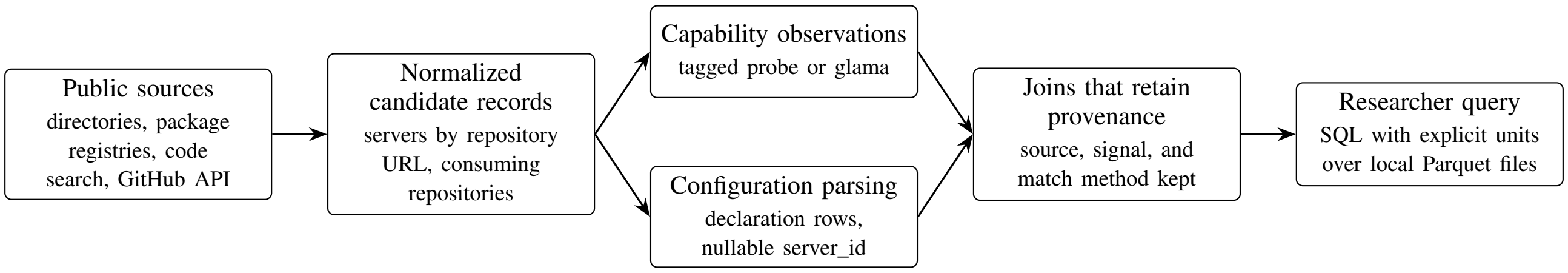
\begin{figure*}[!t]
\centering
\begin{tikzpicture}[
  node distance=3mm and 6mm,
  box/.style={draw,rounded corners=2pt,align=center,font=\footnotesize,
    text width=2.7cm,inner sep=3pt,minimum height=10mm},
  arr/.style={-{Stealth[length=2mm]},semithick}]
\node[box] (src) {Public sources\\[1pt]\scriptsize directories, package registries, code search, GitHub API};
\node[box,right=of src] (cand) {Normalized candidate records\\[1pt]\scriptsize servers by repository URL, consuming repositories};
\node[box,right=of cand,yshift=9mm] (cap) {Capability observations\\[1pt]\scriptsize tagged probe or glama};
\node[box,right=of cand,yshift=-9mm] (cfg) {Configuration parsing\\[1pt]\scriptsize declaration rows, nullable server\_id};
\node[box,right=of cap,yshift=-9mm] (join) {Joins that retain provenance\\[1pt]\scriptsize source, signal, and match method kept};
\node[box,right=of join] (q) {Researcher query\\[1pt]\scriptsize SQL with explicit units over local Parquet files};
\draw[arr] (src) -- (cand);
\draw[arr] (cand.east) -- (cap.west);
\draw[arr] (cand.east) -- (cfg.west);
\draw[arr] (cap.east) -- (join.west);
\draw[arr] (cfg.east) -- (join.west);
\draw[arr] (join) -- (q);
\end{tikzpicture}
\caption{How MCP-Eco is built and used, from public sources to queries with explicit counting units.}
\label{fig:pipeline}
\end{figure*}

We propose MCP-Eco, a dataset that links candidate server projects to consuming repositories
through parsed configuration declarations. On the server side, it contains 78,302 candidate server projects discovered through five directories, six package registries, and code search, with their 65,233 registry listings, 45,272 launch commands and endpoints, and 662,693 declared dependencies. Their capabilities are stored as 284,845 tool, 20,795 resource, and 8,983 prompt records, each tagged as observed by live probing or retrieved from the Glama directory. On the consumer side, it contains 16,165 candidate consuming repositories with 45,529 discovery signals and 9,200 parsed configuration declarations that link repositories to the servers they declare, with the match method recorded for each resolved link. From the associated repositories, it adds 1.73M pull requests, 620K reviews, 301K review comments, 991K issues, 2.07M issue comments, and 9.34M commits, including consumer-side pull requests matched to coding-agent signatures. Every capability, signal, and declaration keeps its source and match method, so a user can restrict an analysis to probed rows, compare directory claims with observed behavior, or audit a single link. Table~\ref{tab:counts} summarizes the release.

\section{Data Collection}
\label{sec:collection}
A Python crawler collected the data in PostgreSQL and exported
the release to Parquet on 2026-06-23 (UTC).
Fig.~\ref{fig:pipeline} summarizes the collection pipeline.
\rqpartlead{Stage 1: Discovering candidate servers.}
Candidates come from three channels. 
(1)~\emph{Directories.} The crawler queries the APIs of the official MCP registry, Glama, Smithery, PulseMCP. For MCP.so, it reads server listing pages found through sitemaps. It collects server names, description, repository URL, package identifiers, and transport details where available. Each Candidate's directory listing is recorded in  \textsf{server\_\allowbreak registry\_\allowbreak listings}. 
(2)~\emph{Package registries.} The crawler searches npm keywords
for \textsf{mcp}, \textsf{mcp-server}, and
\textsf{model-context-protocol}, and runs a free-text search for
\textsf{mcp-server}. It queries the search APIs of NuGet,
crates.io, and Maven Central with \textsf{mcp}.
For PyPI, it filters package names for MCP patterns.
For Docker Hub, it lists images in the \textsf{mcp/} namespace.
Repository URLs are extracted from package metadata where available.
 (3)~\emph{Code search.} GitHub queries look for \textsf{mcpName}
in \textsf{server.json} and \textsf{startCommand} or
\textsf{runtime} in \textsf{smithery.yaml}. These fields indicate
how a server is published or launched. Sourcegraph queries also
look for server SDK imports and implementation patterns across
languages. Repository URLs are normalized and deduplicated.

Repository URLs are normalized by removing trailing slashes and
\textsf{.git} suffixes.
Candidates with the same normalized URL are merged into one
\textsf{mcp\_\allowbreak servers} row.
The \textsf{discovery\_\allowbreak source} field records the channels
that found each candidate. The frozen release contains 78,302
candidate records.

Available star counts reflect discovery-time metadata.
Package identifiers and connection endpoints are stored in
\textsf{mcp\_\allowbreak server\_\allowbreak transports}.
For GitHub repositories, the crawler fetches dependency files
from the repository root through the GitHub Contents API.
These include \textsf{package.json}, \textsf{pyproject.toml},
\textsf{requirements.txt}, \textsf{Cargo.toml}, \textsf{go.mod},
and \textsf{pom.xml}. Declared dependencies are stored in
\textsf{server\_\allowbreak dependencies}, together with their
version constraints, dependency types, ecosystems, and file paths.

\rqpartlead{Stage 2: Observing capabilities.}
Tools, resources, and prompts are stored as observations with a \textsf{source} column, and uniqueness is enforced per source, so the same capability can appear once as advertised and once as observed. Rows with \textsf{source=glama} copy the lists published by the Glama directory, fetched through each server's capability page (the \textsf{/schema}
subpage of its listing). Rows with \textsf{source=probe} come from live probing. 
For each candidate with a transport record, the prober chose one transport
and preferred a remote endpoint over stdio. Remote endpoints were contacted
directly over streamable HTTP or SSE.
After a successful \textsf{initialize} handshake, the prober
requests \textsf{tools/list}, \textsf{resources/list}, and
\textsf{prompts/list}.

Tool records retain input JSON schemas where available.
Resource records retain URIs, MIME types, and descriptions.
Prompt records retain descriptions and argument metadata.
Missing records can reflect skipped probes, connection failures,
or failed listing requests. They do not establish that a server
exposes no capabilities.

\rqpartlead{Stage 3: Discovering consuming repositories.}
GitHub code search is queried for \textsf{mcpServers} in configuration files such as \textsf{claude\_desktop\_config.json},
\textsf{mcp.json}, and \textsf{cline\_mcp\_settings.json},
and under \textsf{.cursor/}. 

Sourcegraph also searches for client SDK imports and framework adapters such as \textsf{from mcp import ClientSession} and
\textsf{MultiServerMCPClient}, alongside dependency,
transport, and infrastructure patterns. 

Every hit is stored in \textsf{repo\_\allowbreak discovery\_\allowbreak signals} with its category (host configuration, client code, framework adapter, or manifest dependency), the matched path, and the pattern. The 45,529 signals cover 16,165 repositories. 

The \textsf{consumer\_\allowbreak type} label summarizes the
evidence. Parsed declarations or host-configuration signals yield
\textsf{ai\_user}. Client-code or framework-adapter signals yield
\textsf{ai\_builder}. Both kinds yield \textsf{both}. Ambiguous evidence yields \textsf{sdk\_unconfirmed}, while
insufficient evidence yields \textsf{unknown}. Server candidates with consumer evidence are retained and marked
\textsf{is\_also\_server}.

\rqpartlead{Stage 4: Parsing and matching declarations.}
For each consuming repository, the crawler requested twelve fixed
configuration paths, such as \textsf{.mcp.json} and \textsf{.cursor/mcp.json}, from the default branch at fetch time. It parsed the \textsf{mcpServers} object of each JSON file found. Each entry becomes one
\textsf{repo\_\allowbreak server\_\allowbreak usage} row, unique by repository, raw server name, and file path. The row keeps the command, the arguments, and a client type derived from the path. This produced 9{,}200 rows.  Each row is then matched to a candidate server using package identifiers, endpoint URLs, or server names. Of the 9{,}200 rows, 5{,}975 matched, giving
5{,}138 distinct repository-server pairs. Resolved declarations receive a \textsf{server\_id}, with the
matching rule recorded in \textsf{match\_method}.
Unresolved declarations keep a null \textsf{server\_id}.

\rqpartlead{Stage 5: Harvesting development records.}
Development records were collected from GitHub-hosted server
candidates and consuming repositories through the GitHub REST API, using Perceval~\cite{perceval2018}.
On the server side, we collected pull requests,
issues, and comments through the GitHub REST API, keeping items updated since
the MCP launch on 2024-11-25. It targeted the 13{,}966 GitHub-hosted
candidates with at least ten stars at discovery. Reviews and review comments were fetched only for a sample of servers: those declared in a configuration on the consumer side, those Glama marks as official, and a deterministic sample of candidates with stored pull requests. On the consumer side, we collected issues  since the MCP launch date for all 16{,}165 repositories. Coding-agent signatures follow Li et al.~\cite{li2025aidev},
with Jules and Codegen added.
The matched signature is stored in \textsf{relevance\_signal}. Commits were extracted from servers and consuming repositories.
We used PyDriller~\cite{spadini2018pydriller} to extract insertions,
deletions, and files changed.
Commits are keyed globally by \textsf{commit\_\allowbreak hash},
retain the first recorded repository URL, and store author dates.
Issues, pull requests, reviews, and comments carry bot flags
based on GitHub account type \textsf{Bot}. 
In total, 22{,}750 repositories have at least one development record.

\rqpartlead{Export.}
On 2026-06-23 every table was exported from PostgreSQL to one Parquet file. \textsf{schema.sql} records the schema and constraints.

\begin{table}[!t]
\caption{Inventory of 16{,}880{,}899 stored records across 19 tables, with identifiers for joining related records.}
\label{tab:counts}
\centering
\footnotesize
\setlength{\tabcolsep}{3pt}
\begin{tabular}{@{}l r >{\raggedright\arraybackslash}p{2.45cm}@{}}
\toprule
\textbf{Table} & \textbf{Records} & \textbf{Main keys} \\
\midrule
\multicolumn{3}{@{}l}{\emph{Server candidates and capabilities}}\\
\quad\textsf{mcp\_registries} & 7 & \textsf{registry\_id} \\
\quad\textsf{mcp\_servers} & 78{,}302 & \textsf{server\_id} \\
\quad\textsf{server\_registry\_listings} & 65{,}233 & \textsf{server\_id}, \textsf{registry\_id} \\
\quad\textsf{mcp\_server\_transports} & 45{,}272 & \textsf{server\_id} \\
\quad\textsf{mcp\_tools} & 284{,}845 & \textsf{server\_id}, \textsf{source} \\
\quad\textsf{mcp\_resources} & 20{,}795 & \textsf{server\_id}, \textsf{source} \\
\quad\textsf{mcp\_prompts} & 8{,}983 & \textsf{server\_id}, \textsf{source} \\
\quad\textsf{server\_dependencies} & 662{,}693 & \textsf{server\_id} \\
\midrule
\multicolumn{3}{@{}l}{\emph{Consuming repositories}}\\
\quad\textsf{consumer\_repos} & 16{,}165 & \textsf{repo\_id} \\
\quad\textsf{repo\_discovery\_signals} & 45{,}529 & \textsf{repo\_id} \\
\midrule
\multicolumn{3}{@{}l}{\emph{Configuration declarations}}\\
\quad\textsf{repo\_server\_usage} & 9{,}200 & \textsf{repo\_id}, \textsf{server\_id} (nullable) \\
\midrule
\multicolumn{3}{@{}l}{\emph{Associated development activity}}\\
\quad\textsf{commits} & 9{,}336{,}613 & \textsf{commit\_hash}, \textsf{repository\_url} \\
\quad\textsf{server\_pull\_requests} & 1{,}610{,}439 & \textsf{pr\_id}, \textsf{server\_id} \\
\quad\textsf{client\_pull\_requests} & 121{,}161 & \textsf{pr\_id}, \textsf{repo\_id} \\
\quad\textsf{pr\_reviews} & 620{,}151 & \textsf{review\_id}, \textsf{pr\_id} \\
\quad\textsf{pr\_review\_comments} & 300{,}883 & \textsf{review\_id}, \textsf{pr\_id} \\
\quad\textsf{pr\_commit\_bridge} & 597{,}344 & \textsf{pr\_id}, \textsf{commit\_hash} \\
\quad\textsf{ecosystem\_issues} & 991{,}441 & \textsf{issue\_id}, \textsf{server\_id} or \textsf{repo\_id} \\
\quad\textsf{issue\_comments} & 2{,}065{,}843 & \textsf{issue\_id} \\
\midrule
\textbf{Total} & 16{,}880{,}899 & \\
\bottomrule
\end{tabular}
\end{table}

\section{Dataset and Access}\label{sec:dataset}
 
\rqpartlead{Data model.} Table~\ref{tab:counts} groups the 19 tables into four blocks. Server-side tables are linked through \textsf{server\_\allowbreak id}, consumer-side tables off \textsf{repo\_\allowbreak id}, and \textsf{repo\_\allowbreak server\_\allowbreak usage} bridges the two with a nullable \textsf{server\_\allowbreak id}. Development tables are keyed by \textsf{pr\_\allowbreak id}, \textsf{issue\_\allowbreak id}, and \textsf{commit\_\allowbreak hash} and point back to a server or a repository. Four provenance columns travel with the data: \textsf{discovery\_\allowbreak source} on candidates, \textsf{source} (\textsf{probe} or \textsf{glama}) on capability rows, \textsf{match\_\allowbreak method} on declarations, and \textsf{relevance\_\allowbreak signal} on consumer pull requests, plus a bot flag on authors. 
% Section~\ref{sec:limitations} lists the join pitfalls these keys imply.
 
\rqpartlead{Access.} The \href{https://huggingface.co/datasets/mcp-research-alliance/mcp-ecosystem}{Hugging Face dataset page} provides public access under CC BY 4.0: one Parquet file per table under \textsf{parquet/}, the PostgreSQL schema in \textsf{schema.sql}, and \textsf{collection\_\allowbreak metadata.json} with the crawler version, and per-table counts. Analyses should pin the frozen revision shown in Listing~\ref{lst:workflow}. Companion code on GitHub\footnote{\url{https://github.com/khalid-tounoussi-1/mcp-eco-artifact}} downloads and verifies the pinned files and reruns both examples below.
 
\rqpartlead{Example for a software engineer.}
Listing~\ref{lst:workflow} shows an engineer who wants a coding assistant to inspect a GitHub issue,
read the relevant source files, and prepare a pull request.
MCP-Eco records the reference GitHub server's probed
\textsf{get\_issue}, \textsf{get\_file\_contents}, and
\textsf{create\_pull\_request} tools in \textsf{mcp\_tools}.
Their stored schemas describe the required inputs.
Following \textsf{server\_id} to \textsf{repo\_server\_usage}
leads to setup examples, including RedisInsight's
\textsf{mcp.json} declaration of
\textsf{@modelcontextprotocol/server-github}.
The dataset therefore helps the engineer inspect tool interfaces
and consult an existing configuration when planning this workflow.
 
\begin{lstlisting}[style=pysql,language=SQL,float=t,
caption={GitHub MCP server tool schemas and a linked project configuration.},
label=lst:workflow]
SELECT t.tool_name, t.input_json_schema,
  c.repository_url, u.config_file_path,
  u.raw_command, u.raw_args, u.match_method
FROM mcp_tools t
JOIN repo_server_usage u USING (server_id)
JOIN consumer_repos c USING (repo_id)
WHERE t.server_id IN (
  SELECT server_id FROM mcp_server_transports
  WHERE package_identifier =
    '@modelcontextprotocol/server-github')
  AND t.source = 'probe'
  AND t.tool_name IN ('get_issue',
    'get_file_contents', 'create_pull_request')
  AND u.match_method IN ('package', 'package_ver')
  AND lower(c.repository_url) =
    'https://github.com/redis/redisinsight'
  AND u.config_file_path = 'mcp.json'
ORDER BY t.tool_name
\end{lstlisting}

\rqpartlead{Example for a researcher.}
A researcher studies how MCP tools support paper-retrieval
workflows. They start with the arXiv server's recorded
\textsf{search\_papers}, \textsf{download\_paper}, and
\textsf{read\_paper} tools.
Following \textsf{server\_id} to \textsf{ecosystem\_issues}
reveals a historical report describing how asynchronous conversion
interrupted the transition from downloading a paper to reading it.
MCP-Eco brings the tool interfaces and developer report together,
providing a concrete case for studying coordination between tools.
 
\section{Research Opportunities}\label{sec:uses}
 
\rqpartlead{Capability interface design.} Tool, resource, and prompt records support studies of how MCP
capabilities are named, structured, and documented.
Researchers can examine tool input schemas, resource URIs
and MIME types, and prompt arguments alongside their
descriptions to investigate interface and documentation quality.
 
\rqpartlead{Configuration practice.} The 9,200 declarations record which servers projects declare, in which host files, with which launch commands, and in which combinations. Questions include which servers are co-declared, how often packages are version-pinned, and how declarations split between local launches and remote endpoints.
 
\rqpartlead{Maintenance and adoption.} Linked configuration and development records enable studies of the relationship between declared server integration and repository maintenance. Researchers can examine whether servers declared by more repositories exhibit different pull-request acceptance rates, merge times, or issue-response patterns. Such analyses must account for collection coverage and distinguish repository-wide activity from MCP-specific changes.

\rqpartlead{Coding-agent contributions.} 
Agentic consumer pull requests support studies of observed coding-agent activity, including acceptance outcomes and review effort in MCP projects.

\section{Limitations, Validation Status, and Ethics}\label{sec:limitations}
 
\rqpartlead{Coverage.} MCP-Eco is a snapshot of a changing ecosystem, bounded by the sources queried and their search limits. Record counts describe this release, not the MCP population. Probing covers only candidates that could be launched or reached without credentials or local setup, and directory metadata can be stale. Declaration counts measure configured linkage, not runtime use. Consumer-side pull requests include only those matched to an agent signature, so they describe agent activity but not its share of all contributions.
 
\rqpartlead{Reliability.} Candidate servers can include false positives and monorepos with several implementations, and consumer labels rest on code-search signals. The token-based matching rule can resolve a bare argument to the wrong candidate (Section~\ref{sec:dataset}), and can hide real links.
 
\rqpartlead{Validation status.}
We verified file integrity against the pinned release, reproduced the reported table counts, and checked schema constraints against the crawler implementation. These checks establish consistency of the exported artifact but do not measure the semantic accuracy of its contents. The precision of server discovery, consumer classification, and declaration matching remains to be evaluated.
 
\rqpartlead{Ethics and redistribution.} Records come from public registries, public code search, and public repositories, and the release metadata records \textsf{scrub\_\allowbreak pii=false}, so commit author emails, GitHub identities, and public issue and pull request text are present in the data. Users should not use the data to profile individuals. The CC BY 4.0 license covers this release and does not relicense third-party content.
 
\section{Conclusion}\label{sec:conclusion}
MCP-Eco provides a public relational dataset connecting candidate Model Context Protocol server projects, their capabilities, consumer configuration declarations, and repository development records. By bringing these sources together, it enables researchers to examine interface design, configuration practices, and maintenance across the MCP ecosystem. Practitioners can inspect recorded tool schemas and consult existing configurations when planning integrations.

To use the dataset, researchers can start from a server, capability, or consuming repository and follow the recorded links to construct a study-specific subset. Capability sources and declaration-matching methods support explicit filtering, while the accompanying schema and analysis scripts provide a starting point for reproducible queries. Analyses should distinguish configured connections from runtime use and account for collection coverage and matching uncertainty. Future releases will prioritize validating server–consumer links, improving entity resolution, and preserving versioned capability observations to support studies of ecosystem evolution. 
\balance
\bibliographystyle{IEEEtran}
\bibliography{references}

\end{document}